\documentclass[pra, aps, twocolumn, groupedaddress, showpacs, superscriptaddress]{revtex4-2}
\usepackage{amssymb,amsmath,amsthm,color,graphicx,times,graphicx}
\usepackage{hyperref}
\usepackage{subfigure}
\usepackage{times}
\usepackage{bbold}
\usepackage{color}

\newcommand{\bra}[1]{\langle{#1}|}
\newcommand{\ket}[1]{|{#1}\rangle}

\providecommand{\openone}{\leavevmode\hbox{\small1\kern-4.3pt\normalsize1}}
\renewcommand\Re{\mathrm{Re}}

\renewcommand{\Re}{\mathrm{Re}}
\renewcommand{\Im}{\mathrm{Im}}

\theoremstyle{plain}

\theoremstyle{definition}

\usepackage{orcidlink}
\makeatletter
\newsavebox{\@brx}
\newcommand{\llangle}[1][]{\savebox{\@brx}{\(\m@th{#1\langle}\)}%
	\mathopen{\copy\@brx\mkern2mu\kern-0.9\wd\@brx\usebox{\@brx}}}
\newcommand{\rrangle}[1][]{\savebox{\@brx}{\(\m@th{#1\rangle}\)}%
	\mathclose{\copy\@brx\mkern2mu\kern-0.9\wd\@brx\usebox{\@brx}}}
\makeatother
\begin{document}
	\title{Quantifying non-Markovianity via local quantum Fisher information}
	\author{Yassine Dakir \orcidlink{0009-0005-3408-1309}}\affiliation{LPHE-Modeling and Simulation, Faculty of Sciences, Mohammed V University in Rabat, Rabat, Morocco.}
	\author{Abdallah Slaoui \orcidlink{0000-0002-5284-3240}}\email{Corresponding author: abdallah.slaoui@um5s.net.ma}\affiliation{LPHE-Modeling and Simulation, Faculty of Sciences, Mohammed V University in Rabat, Rabat, Morocco.}\affiliation{Centre of Physics and Mathematics, CPM, Faculty of Sciences, Mohammed V University in Rabat, Rabat, Morocco.}
	\author{Lalla Btissam Drissi}\affiliation{LPHE-Modeling and Simulation, Faculty of Sciences, Mohammed V University in Rabat, Rabat, Morocco.}\affiliation{Centre of Physics and Mathematics, CPM, Faculty of Sciences, Mohammed V University in Rabat, Rabat, Morocco.}\affiliation{College of Physical and Chemical Sciences, Hassan II Academy of Sciences and Technology, Rabat, Morocco.}
	\author{Rachid Ahl Laamara}\affiliation{LPHE-Modeling and Simulation, Faculty of Sciences, Mohammed V University in Rabat, Rabat, Morocco.}\affiliation{Centre of Physics and Mathematics, CPM, Faculty of Sciences, Mohammed V University in Rabat, Rabat, Morocco.}
	
\begin{abstract}
Characterizing non-Markovianity in open quantum systems (OQSs) is gaining increasing attention due to its profound implications for quantum information processing. This phenomenon arises from the system's evolution being influenced by its previous interactions with the environment. To better understand these complex dynamics, various measures have been proposed, including those based on divisibility, quantum mutual information, and trace distance. Each of these measures provides a different perspective on the behavior of OQSs. Here, we introduce a novel approach to quantifying non-Markovianity by focusing on metrological non-classical correlations. This approach is based on a discord-like measure of quantum correlations for multi-component quantum systems known as local quantum Fisher information (LQFI), which was introduced in [Phys. Rev. A, {\bf 97} (2018) 032326]. It is defined as the minimization of quantum Fisher information with respect to local observables and measurements. Thereby, we examine some examples to clarify the use of our metric by applying it to three different channels: the phase damping channel, the amplitude damping channel, and the depolarizing channel. In contrast to other approaches, this new metric focuses on correlated bipartite $2\otimes d$ systems, has a clear physical interpretation, and effectively captures the features of non-Markovianity. We demonstrate that the non-Markovian or Markovian evolution of an open correlated bipartite system corresponds to an increase or decrease, respectively, in the quantumness of the quantum state. This quantumness is quantified by the maximum of the measurement-induced Fisher information across all local von Neumann measurements. Compared to measuring non-Markovianity based on local quantum uncertainty—a measure of non-classical correlation of the discord type based on single observables—our results confirm that the nonmonotonic evolution of the LQFI reveals a higher degree of backflow of quantum information and is more robust than the corresponding measure based on quantum uncertainty.

\vspace{0.25cm}
\end{abstract}
\date{\today}
	
\maketitle
\section{Introduction}
Over recent years, quantum mechanics has evolved from being a foundational theory in physics to a driving force behind numerous technological revolutions, bridging the gap between theoretical insights and experimental advancements. These quantum technologies, such as quantum computing and quantum communication, have garnered increasing interest due to their revolutionary potential \cite{Dowling2003,Slaoui2023}. However, these technologies face significant challenges due to the inevitable interactions with the environment that lead to a loss of quantum properties through decoherence phenomena \cite{Breuer2002,Dakir2023,Zurek2003}. Weak coupling between the system and its environment characterises the evolution of these quantum systems \cite{Carmichael2009,SlaouiS2020,Ralph2006}. In such weak couplings, it is often reasonable to assume that they are memoryless, i.e., no information flows back into the system later, such that the system dynamics is easier to handle and describe by the Markov and Born approximations hold, which allows us to obtain the time-local master equation within in the Lindblad form \cite{Gorini1976,Lindblad1976}, while in some cases the Markovian approximation breaks down, such as strong coupling between system and environment and non vanishing initial system environment correlation.  In addition, it has been found that non-Markovianity can lead to a significant variety of physical effects in the dynamics of open quantum systems \cite{Jiang2013,Bylicka2016,Man2019,Lorenzo2015,Kutvonen2015} and can serve as a resource in information theory.  If the dynamics are of a Markovian nature, the loss of information will only occur in one direction flow of information from the system to the environment. Conversely, non-Markovian dynamics presents more interesting phenomena due to the memory effect and has been used in various quantum processes. Thus, the problem of the characterization of our non-Markovian effect quickly became a major topic in the theory of open quantum dynamics \cite{Breuer2009,Slaoui2018,Laine2010,Hou2011,Breuer2016}.\par

It has recently been discovered that, in contrast to Markovian processes, non-Markovian processes can have rather diverse impacts on the decoherence and disentanglement of open systems \cite{Zeng2011,Bellomo2007}. Markovian dynamics was unable to adequately characterize a number of pertinent physical systems, including the quantum optical system \cite{Breuer2002}, quantum dots \cite{Kubota2009}, and color-core spin in semiconductors \cite{Kane1998}. The excitation transfer of a biological system \cite{Chin2010,Ishizaki2009} and certain quantum chemistry issues \cite{Shao2004,Pomyalov2005} must also be seen as non-Markovian processes. These distinct properties lead to extensive applications in quantum information processing. These include the preparation of stable entangled states \cite{Huelga2012}, provision of a quantum resource \cite{Laine2012}, improvement of achievable resolution in quantum metrology \cite{Chin2012}, acceleration of quantum evolution \cite{Xu2014,Deffner2013} and more. Theoretically, several measures and non-Markovian witnesses have been proposed, but they may not all precisely coincide when it comes to detecting non-Markovianity. Currently, we possess multiple closely linked yet conceptually different definitions of non-Markovianity. Among the measures are those grounded in the semi-group property \cite{Siudzinska2021,Wolf2008}, trace distance between quantum states \cite{Wipmann2015,Settimo2022}, fidelity \cite{Rajagopal2010}, divisibility \cite{Rivas2010,Rivas2014,Yuen1973}, quantum Fisher information flow \cite{Lu2010,Song2015,Abiuso2023}, quantum capacity \cite{Bylicka2014}, quantum mutual information \cite{Luo2012}, temporal steering \cite{Chen2016}, local quantum uncertainty \cite{He2014} , relative entropy of coherence \cite{He2017}, information backflow \cite{Bylicka2017}, and deviation from time-translational invariance in dynamical maps \cite{Strasberg2018}. Experimentally, has been realized \cite{Xu2010,XuLi2010,Liu2011,Haase2018,Lu2020}. A universally accepted characterization for quantum non-Markovianity in OQSs is currently unavailable and may not exist.\par

The search for detecting non-Markovianity has produced various measures, some of which we have already cited. These non-Markovian measures are equivalent for single-channel open two-level systems \cite{Luo2012,Addis2013}. However, they may not be exactly equivalent for open two-level multichannel systems \cite{Chruscinski2013,Jiang2013}. A universal non-Markovian definition of open quantum systems remains elusive. In this study, we have presented a simple and mathematically viable approach based on the LQFI for measuring non-Markovianity in OQSs. The LQFI proposed by Kim et al \cite{Kim2018}, which is defined as the minimization of the quantum Fisher information associated with a local observable, indicates the maximum information achievable within one of the subsystems. However, the LQFI has considerable potential as a tool for understanding the effects of quantum correlations beyond entanglement and for improving the precision of quantum estimation protocols \cite{Toth2014}. In addition to the derivation of a new LQFI-based measure, the focus of our extension study is to compare it with the local quantum uncertainty measure for the detection of non-Markovianity. The so-called local quantum uncertainty is a measure of quantum correlations proposed by Girolami et al \cite{Girolami2013}, which quantifies the minimum uncertainty resulting from applying local measurements to a subset of the quantum state using the concept of Wigner-Yanase information \cite{Wigner1963}. In addition to the importance of LQU as a quantitative measure of quantum correlations, He et al.\cite{He2014} used it to derive a new measure to detect non-Markovianity. The LQU-based measure provides a clear physical interpretation and captures non-Markovianity in bipartite systems. The quantum uncertainty of local observables in the other free subsystem increases (decreases) in proportion to a non-Markovian (Markovian) process in the other subsystem. Recognizing that both LQU and LQFI share the crucial characteristic of measuring quantum correlations, as evidenced in numerous studies (e.g., Refs. \cite{Dakir2023,Yurischev2023,Mohamed2022}), we will leverage LQFI and its relationship to LQU within the context of quantum correlations to construct our measure for detecting non-Markovianity in open quantum systems. Our previous work \cite{Slaoui2019} demonstrated the clear superiority of LQFI over LQU in any qubit-qudit system. A fundamental question arises: Can LQFI-based and LQU-based measures be compared in quantifying the degree of non-Markovianity, and which would be more effective in this context?\par

In this paper, An alternative measure of non-Markovianity based on the LQFI is presented and contrasted with the measure LQU. Focusing on a bipartite system $AB$, where the primary system $A$ is interacting with the environment and $B$ is the ancilla, we assess the LQFI by applying local unit operations to the ancilla, effectively using it as a measuring device.  In this context, the non-Markovianity of the system evolution is quantified by evaluating the non-monotonic tendencies of the LQFI. Importantly, our measure operates within the framework of completely positive and trace-preserving (CPTP) dynamics, as it relies on the inclusion of an ancillary system, which inherently ties it to the CP-divisibility paradigm (see Fig.4 of Ref.\cite{Breuer2016} for details). This contrasts with measures based solely on positive (P)-divisibility, such as those based on trace distance \cite{Breuer2009}, which are necessary and sufficient to detect non-Markovianity in P-indivisible channels but do not require an ancillary extension. Although P- and CP-divisibility represent distinct approaches to characterizing non-Markovianity, with P-divisibility capturing information backflow in isolated systems and CP-divisibility accounting for correlations in extended systems, our study focuses exclusively on the latter. By focusing on CPTP dynamics, we ensure consistency with ancilla-assisted metrics and maintain a physically meaningful description of open quantum system evolution. Furthermore, it is imperative to highlight that the measure of non-Markovianity derived by the LQFI shows a qualitative consistency with several other established measures used in the context of OQS. In our investigation, we have shown that this LQFI-based measure is consistent with metrics based on the divisibility concept, relative entropy and quantum mutual information and particular LQU. Notably, this alignment holds for CPTP dynamics, as our ancilla-dependent method naturally aligns with CP-divisibility conditions.\par

The structure of this manuscript unfolds as follows:  Section \ref{sec2} begins with a review of the concept of LQFI and local quantum uncertainty. Following this, a non-Markovianity measure based on LQFI for OQSs is introduced. Moving on to section \ref{sec3}, we present illustrative examples to demonstrate the practical utility of this measure, juxtaposing it with the LQU-based measure. Finally, our findings are summarized in section \ref{sec4}.
	
\section{Measuring non-Markovianity}\label{sec2}
In this section, we review fundamental concepts and key equations that will be utilized in the following sections. We adopt an alternative method for measuring and describing non-Markovianity in open quantum systems, based on local quantum uncertainty as proposed by He et al.\cite{He2014}. Then, we present our novel measure based on LQFI and compare it with the LQU-based measure.
\subsection{Non-Markovianity via local quantum uncertainty}
Local quantum uncertainty (LQU) refers to the fundamental unpredictability of quantum phenomena within a local region of a quantum system. This reflects the intrinsically probabilistic nature of quantum behavior, and also highlights the limitations on the precision of simultaneous measurements of certain properties of quantum particles. This discord-type measure of quantum correlations $\mathcal{U}$, proposed by Girolami et al.\cite{Girolami2013}, is defined as the minimum skew information \cite{Wigner1963} from the perspective of local measurements, in any bipartite quantum state. It is explicitly defined as
\begin{equation}
{\cal U}\left(\varrho^{AB}\right)= \min_{L_{A}} {\cal I}\left(\varrho^{AB},L_{A} \otimes  I_{B}\right), \label{LQU}
\end{equation}
where ${\cal I}\left(\varrho^{AB},L_{A} \otimes  I_{B}\right)= -\frac{1}{2}Tr\left[ \sqrt{\varrho^{AB}},L_{A} \otimes  I_{B}\right]$ is the skew information of a density operator $\varrho^{AB}$ with a local observable $L_{A}$. This local observable in its typical form is represented as $L_{A}=\vec{\sigma}.\vec{v}$, with $\arrowvert \vec{v} \arrowvert=1$ and $\vec{\sigma}=\left( \sigma_{1},\sigma_{2},\sigma_{3}\right)$ the Pauli matrices. For any bipartite $2 \otimes d$ system, a closed-form expression for the LQU can be derived as
\begin{equation}
	{\cal U}\left(\varrho^{AB}\right)=1-\delta_{\max}\left({\cal \mathcal{B}}\right),\label{lqu}
\end{equation}
where $\delta_{\max}$ is the largest eigenvalue of the $3\times3$ symmetric matrix ${\cal B}$, whose components are calculated using the following formula
\begin{equation}
	{\cal B}_{ij}={\rm Tr}\left\lbrace \sqrt{\varrho^{AB}}\left( \sigma_{i}\otimes I_{B}\right) \sqrt{\varrho^{AB}}\left(\sigma_{j}\otimes I_{B}\right)\right\rbrace, \label{bij}
\end{equation}
with $\sigma_{i,j}$ $(i,j=1,2,3)$ are the standard Pauli matrices acting on the part $A$. As introduced in Ref.\cite{He2014}, the monotonic decrease of the local quantum uncertainty under local operations in a Markovian process is used to measure non-Markovianity. Deviations from this monotonicity are captured as
\begin{equation}
	\mathcal{N}^{LQU}\left( \Phi\right) =\max_{\varrho^{AB}\left( 0\right)} \int_{\mathcal{D}_{\mathcal{U}}\left( t\right)>0} \mathcal{D}_{\mathcal{U}}\left( t\right) dt,  \label{NLQU}
\end{equation}
where,\\
$\mathcal{D}_{\mathcal{U}}\left( t\right)=\frac{d\mathcal{U}\left( \varrho^{AB}\left( t\right) \right) }{dt}$ with $\varrho^{AB}\left( t\right) =\left(I^{A}\otimes\Phi_{t}^{B}\right) \varrho^{AB}\left( 0\right)$.

\subsection{Non-Markovianity via local quantum Fisher information}
Local quantum Fisher information is a fundamental concept in quantum information theory and quantum estimation theory. Recently, it has emerged as an important measure of non-classical correlations. For a given quantum state $\varrho^{AB}= \sum_{m}h_{m}\ket{\phi_{m}}\bra{\phi_{m}}$ with $h_{m}\geq 1$ and $\sum_{m}h_{m}=1$, the LQFI associated with the local evolution generated by $L_{A}\otimes I_{B}$ can be expressed as \cite{Kim2018,Slaoui2019}
\begin{equation}
\mathcal{F}\left( \varrho^{AB},L_{A}\right)  = \frac{1}{2}\sum_{\substack{m \ne n;\\h_{m}+h_{n}>0}} \frac{\left( h_{m}-h_{n}\right) ^{2}}{h_{m}+h_{n}}\arrowvert\bra{\phi_{m}}L_{A} \otimes  I_{B}\ket{\phi_{n}}\arrowvert^{2}.
\end{equation}
The expression above can be equivalently reformulated as
{\small\begin{equation}
   \mathcal{F}\left( \varrho^{AB},L_{A}\right)  =Tr\left\lbrace  \varrho_{AB} L_{A}^{2}\right\rbrace - \sum_{\substack{m \ne n;\\h_{m}+h_{n}>0}} \frac{2h_{m}h_{n}}{h_{m}+h_{n}}\arrowvert\bra{\phi_{m}}L_{A}\ket{\phi_{n}}\arrowvert^{2},  \label{eq1}
\end{equation}}
Local quantum Fisher information is a fundamental concept in quantum information theory and quantum estimation theory. Recently, it has emerged as an important quantifier of quantum correlations. It is defined as the minimum quantum Fisher information overall local Hamiltonian $L_{A}$ acting in the subspace of party $A$ of the bipartite system $AB$, i.e., 
\begin{equation}
	\mathcal{Q}\left( \varrho^{AB}\right) = \underset{L_{A}}{\text{min}}\mathcal{F}\left( \varrho^{AB}, L_{A}\right), 
\end{equation} 
thus have $Tr\left( \varrho^{AB}, L_{A}^{2}\right)=1$, and the second term in the Eq.(\ref{eq1}) can be expressed as
\begin{equation}
	\sum_{\substack{m \ne n;\\h_{m}+h_{n}>0}} \frac{2h_{m}h_{n}}{h_{m} + h_{n}}\arrowvert\bra{\phi_{m}}L_{A}\ket{\phi_{n}}\arrowvert^{2} =\vec{v}^{\dagger}.\mathcal{S}.\vec{v},
\end{equation}
where the components of the symmetric matrix $\mathcal{S}$ are given by
\begin{equation}
	\mathcal{S}_{kl} = \sum_{\substack{m \ne n;\\h_{m}+h_{n}>0}} \sum\limits_{k,l=1}^{3}\frac{2h_{m}h_{n}}{h_{m} + h_{n}} \bra{\phi_{m}}\sigma_{k}\otimes I\ket{\phi_{n}}\bra{\phi_{n}}\sigma_{l}\otimes I\ket{\phi_{m}} \label{Mkl}.
\end{equation}
To minimize $\mathcal{F}$, it's essential to maximize the expression $\vec{v}^{\dagger}.\mathcal{S}.\vec{v}$ over all unit vectors. This maximum value corresponds to the maximum eigenvalue of $\mathcal{S}$. Therefore, the minimum value of the LQFI is
\begin{equation}
	\mathcal{Q}\left( \varrho^{AB}\right) =1-\eta_{\max}\left( \mathcal{S}\right),
\end{equation}
where $\eta_{\max}\left( \mathcal{S}\right)$ is the maximum eigenvalue of the real symmetric matrix $\mathcal{S}$. LQFI has several properties that make it a valuable measure of quantum correlations \cite{Kim2018}:\par
First, the $\mathcal{Q}\left( \varrho^{AB}\right)$ is preserved under unitary transformations $U$, i.e., $\mathcal{Q}\left(\left( U^{A}\otimes U^{B}\right) \varrho^{AB}\left( U^{A}\otimes U^{B}\right)^{\dagger}\right) =  \mathcal{Q}\left( \varrho^{AB}\right)$.\par
Second, $\mathcal{Q}\left( \varrho^{AB}\right)$ is non-increasing (contractive) under any local completely positive trace preserving (l-CPTP) map $\Phi$ on party, i.e. $Q\left( \varrho^{AB}\right) \geq Q\left(\left(  \Phi^{A} \otimes I^{B}\right) \varrho^{AB}\right)$.\par
By a simple demonstration, we show the extension of the last property on composite systems, we have $\Phi^{A}\left( \varrho^{A}\right) =\sum\limits_{n}\lambda_{n} U_{n}\varrho^{A}U_{n}^{\dagger}$ is a random unitary channel. $U_{n}$ are unitary operators on $L_{A}$, and $\lambda_{n}$ satisfy $\sum_{n}\lambda_{n}=1$, $0\leq \lambda_{n} \leq 1$. Then we have
	\begin{equation}
	\mathcal{F}\left( \left( \Phi^{A} \otimes I^{B}\right) \varrho^{AB},L_{A}\right)\leq \mathcal{F}\left( \varrho^{AB},L_{A}\right).
	\end{equation}
	So,
	\begin{align}
	\mathcal{Q}\left( \varrho^{AB}\right) &= \min_{L_{A}}\mathcal{F}\left( \varrho^{AB},L_{A}\right) \notag\\
	&\geq 	\min_{L_{A}} \mathcal{F}\left( \left( \Phi^{A} \otimes I^{B}\right) \varrho^{AB},L_{A}\right) \notag\\
	&=\mathcal{Q}\left( \left( \Phi^{A} \otimes I^{B}\right) \varrho^{AB}\right).
\end{align}
The last property of LQFI (monotonicity) is paramount in underpinning the formulation of the novel measure of non-Markovianity proposed. This foundational principle is intimately linked with the concept of divisibility . The evolution of a quantum dynamical map, denoted as $\left\lbrace \Phi\right\rbrace$, over time. A quantum dynamical map $\left\lbrace \Phi\right\rbrace$ is divisible if and only if it meets
\begin{equation}
	\Phi_{t}=\Phi_{t,v}\Phi_{v}, \label{divisibility}
\end{equation}
with $\Phi_{t,v}$ signifies completely positive map for any $t >v\geq 0$. In refs \cite{Breuer2009,Laine2010,Hou2011}, all divisible dynamics are defined as Markovian, while non-divisible dynamics are defined as non-Markovian.\par

For Markovian dynamics, the evolution is described by a divisible l-CPTP map $\Phi$. The composite dynamics of the entire system $\varrho^{AB}\left( t\right)$ is given by the Jamiolkowski-Choi isomorphism \cite{Jamiolkowski1972}
\begin{equation}
    \varrho^{AB}\left( t\right) = \left( I^{A}\otimes \Phi_{t}^{B}\right)\varrho^{AB},
\end{equation}
where $I^{A}$ represents the identity operation on the ancillary subsystem A. We have,
\begin{align}
	\mathcal{Q}\left( \varrho^{AB}\left( t\right)\right)&=\mathcal{Q}\left( \left( I^{A}\otimes \Phi_{t}\right) \varrho^{AB}\right) \notag\\
	&= \mathcal{Q}\left( \left( I^{A}\otimes \Phi_{t,v}\Phi_{v}\right) \varrho^{AB}\right) \notag\\
	&=\mathcal{Q}\left( \left( I^{A}\otimes \Phi_{t,v}\right) \varrho^{AB}\left( v\right)\right) \notag\\
	&\leq \mathcal{Q}\left( \varrho^{AB}\left( v\right)\right).
\end{align}
The monotonicity of the LQFI under local operations lays the groundwork for understanding the dynamics of quantum systems. Consequently, the LQFI, $\mathcal{Q}\left( \varrho^{AB}\left( t\right) \right)$ is non-increasing monotone with increasing time.  Hence, which implies that $\frac{d\mathcal{Q}\left( \varrho^{AB}\left( t\right) \right)}{dt} \leq 0$ for any Markovian dynamics \cite{Luo2012,He2014}. In other words, the appearance of non-Markovianity of a dynamical process is implied by the breaking of this monotonicity, i.e.,
	\begin{equation}
	\frac{d\mathcal{Q}\left( \varrho^{AB}\left( t\right) \right)}{dt} > 0.
	\end{equation}
Consequently, we develop a measure of non-Markovianity
\begin{equation}
	\mathcal{N}^{LQFI}\left( \Phi\right) =\max_{\varrho^{AB}\left( 0\right)} \int_{\mathcal{D}_{\mathcal{Q}}\left( t\right)>0} \mathcal{D}_{\mathcal{Q}}\left( t\right) dt,  \label{NLQFI}
\end{equation}
where $\mathcal{D}_{\mathcal{Q}}\left( t\right)=\frac{d\mathcal{Q}\left( \varrho^{AB}\left( t\right) \right) }{dt}$ with $\varrho^{AB}\left( t\right) =\left(I^{A}\otimes\Phi_{t}^{B}\right) \varrho^{AB}\left( 0\right)$, with $\varrho^{AB}\left( 0\right)$ the maximization is performed on all possible initial states. In this context, the above Eq.(\ref{NLQFI}) can also be written as
\begin{equation}
	\mathcal{N}^{LQFI}\left( \Phi\right) =\max_{\varrho^{AB}\left( 0\right) }\sum\limits_{k}\left[ \mathcal{Q}\left( x_{f}^{k}\right)-\mathcal{Q}\left( x_{i}^{k}\right)\right], 
\end{equation}
where the time intervals $\left[ x_{i}^{k} ,x_{f}^{k}\right]$ represent all regions where $\mathcal{D}_{\mathcal{Q}}\left( t\right)>0$, indicating deviations from non-Markovian dynamics. To accurately compute this measure, it is imperative to determine the cumulative increase of the LQFI within each time interval $\left[x_{i}^{k} ,x_{f}^{k}\right]$ and then aggregate the contributions from all such intervals.\par
\vspace{0.5cm}

The LQFI connection to quantum metrology further enhances its relevance, as it is directly linked to the precision of parameter estimation in quantum systems. This makes LQFI particularly useful in scenarios where non-Markovianity impacts measurement precision. Furthermore, LQFI offers greater mathematical flexibility, enabling it to be adapted to various types of interactions and dynamics, making it a more versatile tool across different scenarios. In contrast, LQU may not provide the same level of precision or adaptability, especially in systems with complex dynamics. 
In quantum metrology, quantum Fisher information and skew information satisfy the following inequality relation \cite{Slaoui2019}
	\begin{equation}
	\mathcal{I}\left( \varrho,L\right)  \leq \mathcal{F}\left( \varrho,L\right) \leq 2\mathcal{I}\left( \varrho,L\right),
	\end{equation}
	where
	{\small\begin{equation}
	\mathcal{F}\left( \varrho,L\right) = \frac{1}{2}\sum\limits_{m,n}\left( 1+\frac{2\sqrt{h_{m}h_{n}}}{h_{m}+h_{n}}\right)\left( \sqrt{h_{m}}-\sqrt{h_{n}}\right)^{2}\arrowvert\bra{\phi_{m}}L\ket{\phi_{n}}\arrowvert^{2},
	\end{equation}}
	\begin{equation}
	\mathcal{I}\left( \varrho,L\right) = \frac{1}{2}\sum\limits_{m,n}\left( \sqrt{h_{m}}-\sqrt{h_{n}}\right)^{2}\arrowvert\bra{\phi_{m}}L\ket{\phi_{n}}\arrowvert^{2}.
	\end{equation}
	Correspondingly, the derivatives of the QFI $\mathcal{F}\left( \varrho,L\right)$ can be obtained as
\begin{align}
	\frac{d\mathcal{F}\left( \varrho,L\right)}{dt}=\sum_{m,n}\Bigg[ &\left( 1+\frac{2\sqrt{h_{m}h_{n}}}{h_{m}+h_{n}}\right)\frac{d\mathcal{I}\left( \varrho,L\right)}{dt} \notag\\
	&+\mathcal{I}\left( \varrho,L\right)\frac{d}{dt}\left( 1+\frac{2\sqrt{h_{m}h_{n}}}{h_{m}+h_{n}}\right)\Bigg], \label{fisher}
\end{align}
where
\begin{equation*}
	\frac{d}{dt}\left( 1+\frac{2\sqrt{h_{m}h_{n}}}{p_{m}+p_{n}}\right)=\frac{\dot{h}_{m}h_{n}\left( 1-2h_{m}\right)+h_{m}\dot{h}_{n}\left( 1-2h_{n}\right) }{\left( h_{m}+h_{n}\right)^{2}\sqrt{h_{m}h_{n}}},
\end{equation*}
with $0\leq h_{m,n}\leq 1$ and $0\leq \dot{h}_{m,n}\leq 1$, we find that
\begin{equation}
	0\leq \frac{d}{dt}\left( 1+\frac{2\sqrt{h_{m}h_{n}}}{h_{m}+h_{n}}\right)\leq \frac{1}{2}.  \label{en1}
\end{equation}
By of Eq. (\ref{fisher}) and Eq. (\ref{en1}) we obtain the new inequality relation
\begin{equation}
	\frac{d\mathcal{I}\left( \varrho,L\right)}{dt} \leq 	\frac{d\mathcal{F}\left( \varrho,L\right)}{dt}\leq 	2\frac{d\mathcal{I}\left( \varrho,L\right)}{dt} \label{dr}
\end{equation}
Due to the relationship between the derivatives of the skew information and the derivatives of the quantum Fisher information (\ref{dr}), the above inequality relation is reduced to
\begin{equation}
	\frac{d\mathcal{U}\left( \varrho\right)}{dt} \leq 	\frac{d\mathcal{Q}\left( \varrho\right)}{dt}\leq 	2\frac{d\mathcal{U}\left( \varrho\right)}{dt}
\end{equation}
A regime is classified as non-Markovian if $\frac{d \mathcal{X} \left( \varrho\right)}{dt}>0$, with $\mathcal{X}\equiv \left( \mathcal{U},\mathcal{Q}\right)$. 
\begin{equation}
	\mathcal{N}^{LQU}\leq \mathcal{N}^{LQFI}\leq 2\mathcal{N}^{LQU} \label{egalite}
\end{equation} 
\section{Applications} \label{sec3}
In this section, we explore several applications of our measure for evaluating non-Markovian processes. By examining these examples, we aim to elucidate the effectiveness of our measure in characterizing quantum dynamical systems.
\subsection{Phase damping channel} \label{part1}
In this section, our primary focus revolves around the examination and evaluation of the proposed non-Markovianity procedure. Specifically, our objective is to contrast the measure of non-Markovianity based on LQU with our measure (LQFI-based). To provide a concrete context for this analysis, we delve into the typical model depicting a single-qubit dephasing channel. To further elucidate this model, we present the Hamiltonian governing the interaction between a single-qubit and a thermal reservoir \cite{Breuer2002}
\begin{equation}
    H=\omega_{0}\sigma_{z}+\sum\limits_{i}\omega_{i}a_{i}a_{i}^{\dagger}+\sum\limits_{i}\left( g_{i}\sigma_{z}a_{i}+g_{i}^{*}\sigma_{z}a_{i}^{\dagger}\right),
\end{equation}
where $\omega_{0}$ denotes the qubit resonant transition frequency, while $a_{i}\left( a_{i}^{\dagger}\right)$ represents the annihilation (creation) operators, The frequency of the ith reservoir mode is denoted by $\omega_{i}$, and $g_{i}$ signifies the reservoir-qubit coupling constant associated with each mode. We consider the scenario where a qubit is subjected to non-Markovian phase-damping noise. The dynamic evolution of the system can be succinctly expressed in the form of the master equation
\begin{equation}
	\dot{\varrho}\left( t\right) = \gamma\left( t\right) \left( \sigma_{z} \varrho\left( t\right) \sigma_{z}-\varrho\left( t\right) \right). \label{master1}
\end{equation}
The dynamical map that governs the dephasing channel acting on a single-qubit system described by the density operator can be represented as follows
\begin{equation}
	\varrho\left(t\right)=\Phi_{t}\varrho\left( 0\right) =\begin{pmatrix}
	a  &&  P\left(t\right) b
	\\ P\left(t\right) c && d
	\end{pmatrix},
\end{equation}
where $P\left(t\right) =\exp\left( -2\Lambda\left(t\right)\right)$ with $\Lambda\left(t\right) =\int_{0}^{t} \gamma\left( v\right) dv$, the decoherence rate is given by the relation for a zero-temperature reservoir with spectral density $J\left(\omega\right)$ \cite{Breuer2002,Haikka2013}
\begin{equation}
	\gamma\left( t\right) = \int  J\left(\omega\right)\frac{\sin\left( \omega t\right)}{\omega}d\omega.
\end{equation}
For initial state $\varrho^{AB}\left( 0\right) = \ket{\psi}^{AB}\bra{\psi}$, the total density operator is
\begin{equation}
	\varrho^{AB}\left( t\right) =\left( I^{A}\otimes \Phi_{t}^{B}\right) \varrho^{AB}\left( 0\right)=\frac{1}{2}\begin{pmatrix}
	1 && 0 && 0 && P\left(t\right) 
	\\0 && 0 && 0 && 0
	\\0 && 0 && 0 && 0
	\\P\left(t\right) && 0 && 0 && 1
	\end{pmatrix},
\end{equation}
where $\ket{\psi}=\frac{1}{\sqrt{2}}\left( \ket{00}+\ket{11}\right)$. To compute the explicit expression of the LQFI,one must first calculate the elements of the matrix $\mathcal{S}$. Based on Equation.(\ref{Mkl}), it's verified that all elements are equal to zero except $\mathcal{S}_{33}$, which is written as follows $1-P\left(t\right)^{2}$. Thus, in this scenario, the LQFI is represented as
\begin{equation}
   \mathcal{Q}\left( \varrho^{AB}\left( t\right)\right) =P^{2}\left(t\right),
\end{equation}
and
\begin{equation}
    \mathcal{D}_{\mathcal{Q}}\left( t\right)=-4\gamma\left( t\right) P^{2}\left(t\right) \label{dlq}
\end{equation} 
Having derived the expression for LQFI in Eq.(\ref{dlq}), we have formulated the explicit expression for the non-Markovianity measure $\mathcal{N}^{LQFI}(\Phi_t)$:
\begin{align}
	\mathcal{N}^{LQFI}\left(\Phi_{t}\right)&= -\int_{\mathcal{D}_{\mathcal{Q}}\left( t\right)>0}4\gamma\left( t\right) P^{2}\left(t\right)dt\notag\\
	&= -\int_{\gamma\left( t\right)<0}4\gamma\left( t\right) P^{2}\left(t\right)dt. \label{n22}
\end{align}
Interestingly, $\mathcal{N}^{LQFI}\left(\Phi_{t}\right)>0$ is analogous to $\gamma\left( t\right) <0$, which is consistent with both the measures based on the quantum trace distance, quantum mutual information, and local quantum uncertainty\cite{Luo2012,He2014}. This has been previously emphasized in works \cite{Vacchini2011,Chrus2022}. We consider a reservoir spectral density defined by the expression $J(\omega) = (\omega/\omega_c)^{s}e^{-\omega/\omega_c}$, where $\omega_c$ is the reservoir cutoff frequency and $s$ is the ohmicity parameter. The parameter $s$ allows for the transition between different reservoir types: sub-Ohmic for $s < 1$, Ohmic for $s = 1$, and super-Ohmic for $s > 1$. An analytical expression can be found for $\gamma\left( t\right)$ at zero temperature
\begin{equation}
\gamma\left( t\right) =\frac{\omega_{c}\Gamma\left[ s\right] \sin\left( \arctan\left( \omega_{c}t\right)\right)}{\left( 1+\left( \omega_{c}t\right)^{2}\right)^{s/2} },
\end{equation}
where $\Gamma\left[ x\right]$ the Euler gamma function.
\begin{widetext}

\begin{figure}[hbtp]
			{{\begin{minipage}[b]{.5\linewidth}
						\centering
						\includegraphics[scale=0.435]{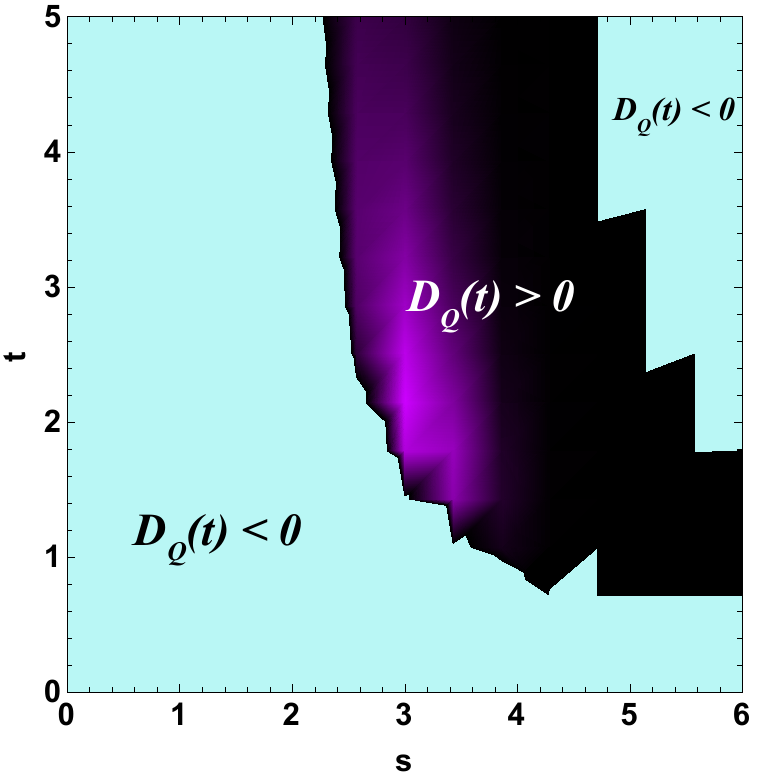} \vfill $\left(a\right)$
					\end{minipage}\hfill
					\begin{minipage}[b]{.5\linewidth}
						\centering
						\includegraphics[scale=1.3]{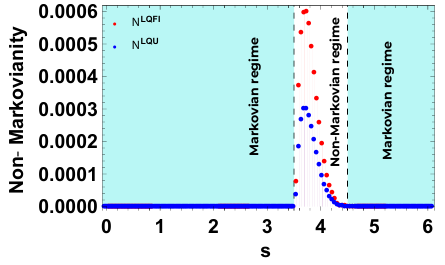} \vfill \vfill  $\left(b\right)$
			\end{minipage}}}
			\caption{(a) The behaviors of $\mathcal{D}_{\mathcal{Q}}\left( t\right)$ (\ref{dlq}) as a function of time $t$ and ohmicity parameter $s$ (b) the non-Markovianity $\mathcal{N}^{LQFI}$, $\mathcal{N}^{LQU}$ as a function of the $s$ for a phase damping channel.}\label{Fig5}
\end{figure}
\end{widetext}
In Fig. \ref{Fig5} our study aims to investigate the behavior of $\mathcal{D}_{\mathcal{Q}}\left( t\right)$ and non-Markovianity for a phase damping channel.

The LQFI increases for those times when $\gamma\left( t\right)<0$, meaning we will obtain $\mathcal{D}_{\mathcal{Q}}\left( t\right)>0$ during these intervals this situation is demonstrated in Fig.(\ref{Fig5}a). The aspect of negativity indicates both a violation of the divisibility property (\ref{divisibility}) and a violation of monotonicity. In Fig.(\ref{Fig5}(b)) we compare our measure $\mathcal{N}^{LQFI}$ (red line) with another measure based on LQU to quantify non-Markovianity $\mathcal{N}^{LQU}$ (blue line) with an ohmic spectral density. The dashed region corresponding to $\mathcal{D}_{\mathcal{Q}}>0$ represents the Markovian regime, i.e., the light blue colored region as shown in Fig.(\ref{Fig5}(b)). We find that the non-Markovian regime exists in the super-ohmic region where $s>3.5$. We find that the non-Markovian regime exists in the super-ohmic region where $s>3.5$. The behavior of non-Markovianity measured by $\mathcal{N}^{LQFI}$ and measured by $\mathcal{N}^{LQU}$ is remarkably similar, as both show an initial increase, followed by a subsequent decrease as $s$ increases, and finally disappears when $s<4.5$. Furthermore, it is clear from the results shown in Fig.(\ref{Fig5}b) that the degree of non-Markovianity measured by the LQFI-based measure exceeds that obtained by the LQU-based measure. This observation is consistent with the inequality (\ref{egalite}), which confirms that the degree of non-Markovianity measured by $\mathcal{N}^{LQFI}$ exceeds that measured by $\mathcal{N}^{LQU}$.\par
\subsection{Amplitude damping channel}
Here, we demonstrate how our measure could be calculated through a typical quantum process. We consider on typical quantum scenario where system $A$ is initially correlated with device B, represented by a two-level (qubit) system \cite{He2011}. Additionally, the apparatus is in contact with a zero-temperature reservoir, with the primary effect being the introduction of an amplitude damping process solely on the apparatus $A$. This is a phenomenon that can be aptly described by the following Hamiltonian expression \cite{Breuer2002}.
\begin{equation}
H=\omega_{0}\sigma_{+}\sigma_{-}+\sum\limits_{i}\omega_{i}b_{i}^{\dagger}b_{i}+\sum\limits_{i}\left(g_{i}b_{i}\sigma_{+}+g_{i}^{*}b_{i}^{\dagger}\sigma_{-}\right) 
\end{equation}
where $\sigma_{+}\left( \sigma_{-}\right)$ represents the raising (lowering) Pauli operator. It's associated dynamics is described using a master equation is
\begin{align}
	\dot{\varrho}\left( t\right)=-&\frac{i}{2}h\left( t\right) \left[ \sigma_{+}\sigma_{-},\varrho\left( t\right) \right] +\notag\\& \gamma\left( t\right) \left[ \sigma_{-}\varrho\left( t\right) \sigma_{+}-\frac{1}{2}\left\lbrace \sigma_{+}\sigma_{-},\varrho\left( t\right) \right\rbrace \right],
\end{align}
with 
\begin{align}
	&h\left( t\right)=-2\Im\frac{\dot{\mathfrak{R}}\left( t\right)}{\mathfrak{R}\left( t\right)},\\&\gamma\left( t\right) =-2\Re\frac{\dot{\mathfrak{R}}\left( t\right)}{\mathfrak{R}\left( t\right)}=-\frac{2}{\arrowvert \mathfrak{R}\left( t\right) \arrowvert}\frac{d}{dt}\arrowvert \mathfrak{R}\left( t\right) \arrowvert,
\end{align}
where $\mathfrak{R}\left( t\right)$ check the following integro-differential equation with the initial condition $\mathfrak{R}\left( 0\right)=1$, 
	\begin{equation}
	\dot{\mathfrak{R}}\left( t\right) = -\int_{0}^{t}g\left( t-t_{1}\right) \mathfrak{R}\left( t_{1}\right) dt_{1} \label{R}
	\end{equation}
	where the two-point reservoir correlation function $g\left( t-t_{1}\right)$ is derived from the Fourier transform of the reservoir spectral density;
	\begin{equation}
	g\left( t-t_{1}\right)=\int  J\left( \omega\right) \exp\left[ i\left( \omega_{0}-\omega\right) \left( t-t_{1}\right)\right]d\omega. \label{f}
	\end{equation}
	The evolution of the system qubit  $\varrho\left( t\right)$ is determined by the dynamical map is given by
	\begin{equation}
	\varrho\left( t\right) =\Phi\left( t\right) \varrho\left( 0\right)= \begin{pmatrix}
	1-\arrowvert \mathfrak{R}\left( t\right) \arrowvert^{2}d   && \mathfrak{R}^{*}\left( t\right)b 
	\\\mathfrak{R}\left( t\right)c && \arrowvert \mathfrak{R}\left( t\right) \arrowvert^{2}d
	\end{pmatrix}.
	\end{equation}
We examine only the case where the reservoir spectral density has a Lorentzian distribution, i.e,
$J\left( \omega\right) =\gamma_{0}\lambda^{2}/2\pi\left( \left( \omega-\omega_{c}\right)^{2}+\lambda^{2}\right)$ \cite{Breuer2002},
where $\omega_{c}$ is the central frequency of the distribution, $\lambda$ is the width of the Lorentzian, and $\gamma_{0}$ is the system-reservoir coupling constant. By solving Eq. (\ref{R}) with the correlation function, we derive the following \cite{Maniscalco2006}
\begin{equation}
	\mathfrak{R}\left( t\right)=e^{\frac{-\left( \lambda-i\Delta\right)t}{2}}\left[  \cosh\left( \frac{\Omega t}{2}\right) +\frac{\lambda-i\Delta}{2}\sinh\left( \frac{\Omega t}{2}\right) \right],
\end{equation}
where $\Omega=\sqrt{\left( \lambda-i\Delta\right)^{2}-2\gamma_{0}\lambda}$ and $\left( \Delta=\omega_{0}-\omega_{c}\right)$  is the system-reservoir frequency detuning, with $\omega_{0}$ the qubit frequency.\\

Now, considering the effect of a single-qubit amplitude damping, the resulting time-evolved density matrix can be expressed as:
	\begin{equation}
	\varrho^{AB}\left( t\right)=\frac{1}{2}\begin{pmatrix}
	1 && 0 && 0 && \mathfrak{R}^{*}\left( t\right)  
	\\0 && 1-\arrowvert \mathfrak{R}\left( t\right)\arrowvert^{2} && 0 && 0
	\\0 && 0 && 0 && 0
	\\\mathfrak{R}\left( t\right)  && 0 && 0 && \arrowvert \mathfrak{R}\left( t\right)\arrowvert^{2}
	\end{pmatrix} \label{rho3}
	\end{equation}
By a simple arithmetic operation, we find that the expression of the LQFI in this system is written as follows
\begin{equation}
	\mathcal{S}_{11}=\mathcal{S}_{22}=1-\arrowvert \mathfrak{R}\left( t\right)\arrowvert^{2},\quad \mathcal{S}_{33}=0
\end{equation}
The final expression of the LQFI in the following form
	\begin{equation}
	\mathcal{Q}\left( \varrho^{AB}\left( t\right) \right)=\arrowvert \mathfrak{R}\left( t\right)\arrowvert^{2},
	\label{Qa}
	\end{equation}
In turn, the derivative of LQFI can be obtained in the following form
	\begin{equation}
	\mathcal{D}_{\mathcal{Q}}\left( t\right)= 2\arrowvert \mathfrak{R}\left( t\right) \arrowvert \frac{d\arrowvert \mathfrak{R}\left( t\right) \arrowvert}{dt}, \label{dq1}
	\end{equation}
Consequently, the measure of non-Markovianity, $\mathcal{N}^{LQFI}\left(\Phi_{t}\right)$ can be expressed as
\begin{equation}
	\mathcal{N}^{LQFI}\left(\Phi_{t}\right) =-\int_{\gamma\left( t\right) <0}  \gamma\left( t\right) \arrowvert \mathfrak{R}\left( t\right) \arrowvert^{2} dt,
\end{equation}
Evidently, the condition $\mathcal{N}^{LQFI}\left( \Phi_{t}\right)>0$  is equivalent to $\frac{d\arrowvert \mathfrak{R}\left( t\right) \arrowvert}{dt}>0$ or $\gamma\left( t\right) <0$. This is consistent with other measures, such as those based on dynamical divisibility, mutual quantum information and local quantum uncertainty of the amplitude damping channel.\\
\begin{widetext}

\begin{figure}[hbtp]
			{{\begin{minipage}[b]{.39\linewidth}
						\centering
						\includegraphics[scale=0.63]{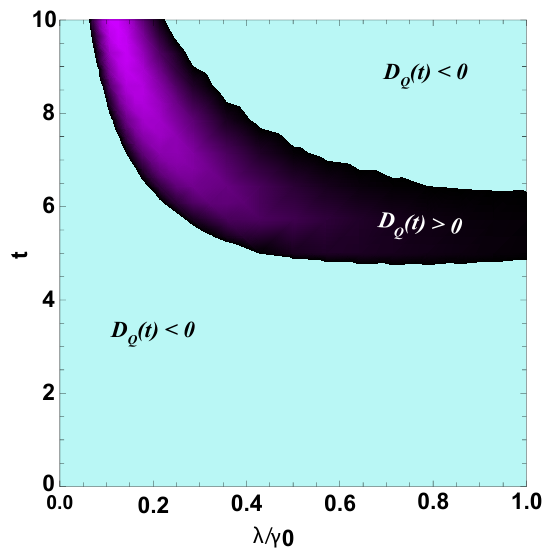} \vfill $\left(a\right)$
					\end{minipage}\hfill
					\begin{minipage}[b]{.5\linewidth}
						\centering
						\includegraphics[scale=1.3]{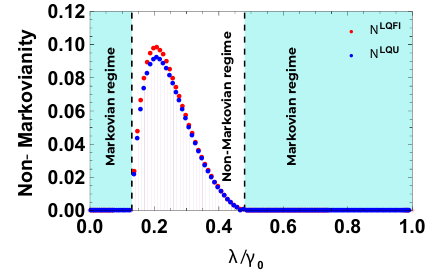} \vfill \vfill  $\left(b\right)$
			\end{minipage}}}
			\caption{(a) The behaviors of $\mathcal{D}_{\mathcal{Q}}\left( t\right)$ (\ref{dq1}) as a function of time $t$ and $\lambda/ \gamma_{0}$. (b) the non-Markovianity $\mathcal{N}^{LQFI}$, $\mathcal{N}^{LQU}$ as a function of the $\lambda/\gamma_{0}$ for the amplitude damping channel}\label{Fig6}
\end{figure}
\end{widetext}

The plot in Fig. (\ref{Fig6}) shows the behavior of the derivative of $\mathcal{D}_{\mathcal{Q}}(t)$ as a function of time $t$ and $\lambda/\gamma_{0}$ (as shown in Fig. (\ref{Fig6}a)), and the dynamics of measure non-Markovianity as a function of $\lambda/\gamma_{0}$ (as shown in Fig. (\ref{Fig6}b)) for the amplitude damping channel. 

For Fig. (\ref{Fig6}a) we provides a comprehensive visualization of the time derivative of $\mathcal{D}_{\mathcal{Q}}(t)$ as a function of time and $\lambda/\gamma_{0}$. Remarkably, the observed positivity in certain time intervals of $\mathcal{D}_{\mathcal{Q}}(t)$ indicates that the dynamics of the system is non-Markovian this is represented by the area colored purple. This deviation from Markovian behavior suggests the presence of memory effects within the system evolution where past states influence future dynamics a feature associated with complex quantum systems. In this context, Fig.(\ref{Fig6}b) shows $\mathcal{N}^{LQFI}$ and $\mathcal{N}^{LQU}$ in the phase damping dynamical model. The non-Markovian nature is clearly evident in the region $0.18 \leq \lambda/\gamma_{0} \leq 0.43$. Interestingly, the degree of non-Markovianity exhibits an increasing behavior as it escalates with increasing values of $\lambda/\gamma_{0}$ until it reaches a critical point where it peaks. Beyond that, there is a gradual decrease until it disappears. These results are particularly intriguing for the comparison between $\mathcal{N}^{LQFI}$ and $\mathcal{N}^{LQU}$ and provide valuable insights into the nature of non-Markovian dynamics. The consistently higher values of $\mathcal{N}^{LQFI}$ compared to $\mathcal{N}^{LQU}$ across different parameter regimes underscore the robustness of the former in capturing the dynamics of non-Markovianity. This observation reinforces the theoretical framework provided by the inequality presented in Eq.(\ref{egalite}).
\subsection{Depolarizing channel}
As a further extension of our methodology, we investigate the non-Markovianity of a non-Markovian depolarizing channel. According to Daffer et al \cite{Daffer2004}, they have studied in detail the dynamical characteristics of this system, particularly the total positivity requirements of the map corresponding to a master equation. In this specific model, we can define the time-dependent Hamiltonian using a two-level system subject to random telegraph noise as follows
\begin{equation}
	H\left( t\right) = \hbar\sum\limits_{i=1}^{3} \Gamma_{i}\left( t\right) \sigma_{i} \label{H3}
\end{equation}
Where $\Gamma_{i}=a_{i}q_{i}\left( t\right)$ be independent random variables and $\sigma_{i}$ be the standard Pauli operators. $q_{i}\left( t\right)$ has a Poisson distribution taking the value $t/2\tau_{i}$, and $a_{i}$ is an independent coin flip random variable taking the value $\pm a_{i}$. The equation of motion for the density operator, corresponding to the time-dependent Hamiltonian given in Eq.(\ref{H3}), is described by the von Neumann equation
\begin{equation}
	\dot{\varrho}\left( t\right) = -i/\hbar \left[ H\left( t\right) , \varrho\right] = -i\sum_{k}\Gamma_{k}\left( t\right) \left[ \sigma_{k},\varrho\right].
\end{equation}
The formal solution for this equation is as follows
\begin{equation}
    \varrho\left( t\right) = \varrho\left( 0\right) -i\int_{0}^{t}\sum_{k}\Gamma_{k}\left( t\right) \left[ \sigma_{k},\varrho\right]ds. \label{r1}
\end{equation}
The following memory kernel master equation can be obtained by replacing the formal solution Eq.(\ref{r1}) with the von Neumann equation, then executing a stochastic average
\begin{equation}
	\dot{\varrho}\left( t\right) = -\int_{0}^{t}\sum_{k}e^{-(t-t^{'})/\tau_{k}^{2}}\left[ \sigma_{k},\left[ \sigma_{k},\varrho\left( t^{'}\right)\right]\right] dt^{'},
\end{equation}
where $<\Gamma_{j}\left( t\right)\Gamma_{k}\left( t^{'}\right)> = b_{k}^{2}\exp\left( -\arrowvert t-t^{'}\arrowvert /\tau_{k}\right) \delta_{jk}$ is the memory kernel comes from the correlation functions of random telegraph signal.\par
As the focus of this study revolves around bipartite systems, it is possible to define the time evolution of initial density operators as \cite{Kraus1983}
\begin{equation}
    \varrho\left( \nu\right) = \sum\limits_{k,l=0}^{3}\left( E_{k} \otimes E_{l}\right)^{\dagger}\varrho\left(0 \right) \left( E_{k} \otimes E_{l}\right)  \label{rhon}
\end{equation}
where dimensionless time $\nu=\frac{t}{2\tau}$ , with $\tau_{i}=\tau$. The Kraus operators $E_{k,l}=\sqrt{p_{k,l}\left( \nu\right)}\sigma_{k,l}$. The non-negative parameters $p_{k,l}\left( \nu\right)$ are defined by 
\begin{align}
	&p_{0}\left( \nu\right) = \frac{1}{4}\left[ 1+\Upsilon_{1}\left( \nu\right) +\Upsilon_{2}\left( \nu\right) +\Upsilon_{3}\left( \nu\right) \right],\\
	&p_{1}\left( \nu\right) = \frac{1}{4}\left[1+\Upsilon_{1}\left( \nu\right) +\Upsilon_{2}\left( \nu\right) -\Upsilon_{3}\left( \nu\right) \right],\\
	&p_{2}\left(\nu\right) = \frac{1}{4}\left[ 1-\Upsilon_{1}\left( \nu\right) +\Upsilon_{2}\left( \nu\right) -\Upsilon_{3}\left(\nu\right) \right],\\
	&p_{3}\left( \nu\right) =  \frac{1}{4}\left[1-\Upsilon_{1}\left(\nu\right)-\Upsilon_{2}\left( \nu\right) +\Upsilon_{3}\left(\nu\right) \right].
\end{align}
where
\begin{equation}
\Upsilon_{i}\left( \nu\right) = e^{-\nu}\left( \cos\left( \mu_{i}\nu\right) +\sin\left( \mu_{i}\nu\right)/\mu_{i}\right),
\end{equation}
with $\mu_{i}=\sqrt{\left( 4\kappa_{i}\tau\right)^{2}-1}$, and $k_{i}^{2}=b_{j}^{2}+b_{k}^{2}$ for $ i \neq j \neq  k,l$.
For simplicity, in this study we examine the case where the noise acts in all directions $x$, $y$ and $z$. Particularly, provided that the condition $b_{1}=b_{2}=b_{3}=b$ thus, $\Upsilon_{1}\left( \nu\right) =\Upsilon_{2}\left( \nu\right) =\Upsilon_{3}\left( \nu\right) \equiv  \Upsilon\left(\nu\right)$, $p_{1}\left( \nu\right)=p_{2}\left( \nu\right)=p_{3}\left( \nu\right)\equiv \frac{1-\Upsilon\left( \nu\right) }{4}$ and $\Upsilon_{0}\left( \nu\right)= \frac{1+3\Upsilon\left( \nu\right) }{4}$.\\
To illustrate our discussion, we will initially assume that both qubits are in a maximally mixed state category, as described by the X-structured density operator \cite{Slaoui2018}.
\begin{equation}
	\varrho\left( 0\right) = \frac{1}{4}\left( I \otimes I+ \sum\limits_{i=1}^{3}r_{i}\sigma_{i}\otimes\sigma_{i}\right) \label{rho0}
\end{equation}
In which the correlation parameters satisfy the conditions $0\leq \arrowvert r_{i}\arrowvert \leq 1$ . For simplicity, we take $\arrowvert r_{1}\arrowvert \geq \arrowvert r_{2} \rvert$. In the computational basis $\left\lbrace\ket{00}, \ket{01},\ket{10}, \ket{11}\right\rbrace $, the density matrix is expressed as
\begin{equation}
	\varrho\left( 0\right) =\frac{1}{4}\begin{pmatrix}
		r_ {3}^{+} && 0 && 0 && r_{-}
		\\0 && r_ {3}^{-} && r_{+} && 0
		\\0 && r_{+} && r_ {3}^{-} && 0
		\\r_{-} && 0 && 0 && r_ {3}^{+}
	\end{pmatrix} 
\end{equation}
where $r_{3}^{+}=1+r_{3}$,  $r_{3}^{-}=1-r_{3}$, $r_{+}=r_{1}+r_{2}$, $r_{-}=r_{1}-r_{2}$. Using initial state of Eq. (\ref{rho0}), from Eq. (\ref{rhon}), we can derive the evolution of the density matrix
\begin{equation}
	\varrho\left( \nu\right) =\begin{pmatrix}
		\vartheta^{+}\left( \nu\right) && 0 && 0 && w^{-}\left( \nu\right)
		\\0 && \vartheta^{-}\left( \nu\right) && w^{+}\left( \nu\right) && 0
		\\0 && w^{+}\left( \nu\right) && \vartheta^{-}\left( \nu\right) && 0
		\\w^{-}\left( \nu\right) && 0 && 0 && \vartheta^{+}\left( \nu\right)
	\end{pmatrix}
\end{equation}
where
$\vartheta^{\pm}\left( \nu\right)=\frac{1}{8}\left( r_{3}^{+}+r_{3}^{-}\pm\left( r_{3}^{+}-r_{3}^{-}\right) \Upsilon^{2}\left( \nu\right) \right)$, \quad $w^{\pm}\left( \nu\right)=\frac{r_{\pm}\Upsilon^{2}\left( \nu\right)}{4}$. In order to determine the explicit expression for LQFI, the initial step involves calculating the elements of the matrix $\mathcal{S}$. When using Eq. (\ref{Mkl}), one gets:
\begin{equation}
	\mathcal{S}_{11}= \frac{1-\Theta }{1-\left( r_{1}\right) ^{2}\Upsilon^{4}\left( \nu\right)}, \quad \mathcal{S}_{22}=\frac{1-\Theta }{1-\left( r_{2}\right) ^{2}\Upsilon^{4}\left( \nu\right)}
\end{equation}
\begin{equation}
	\mathcal{S}_{33}=\frac{1-\Theta }{1-\left( r_{3}\right) ^{2}\Upsilon^{4}\left( \nu\right)}
\end{equation}
where $\Theta=\left( \left( r_{1}\right) ^{2}+\left( r_{2}\right) ^{2}+\left( r_{3}\right) ^{2}\right)\Upsilon^{2}\left( \nu\right)-2r_{1}r_{2}r_{3}\Upsilon^{6}\left( \nu\right)$.\\
We have  $\arrowvert r_{1}\arrowvert \geq \arrowvert r_{2} \arrowvert$. Thus, one has $\mathcal{S}_{11}\geq\mathcal{S}_{22}$,  the final expression of LQFI and its derivative is given
\begin{equation}
	Q\left( \varrho\left( \nu\right) \right) =1-\max\left\lbrace 	\mathcal{S}_{11},	\mathcal{S}_{33}\right\rbrace ,
\end{equation}
\begin{equation}
	\mathcal{D}_{\mathcal{Q}}\left( \nu\right) =-\max\left\lbrace\frac{d\mathcal{S}_{11}}{d\nu},	\frac{d\mathcal{S}_{33}}{d\nu}\right\rbrace.
\end{equation}
	\begin{figure}[ht]
		\centering
		\includegraphics[width=0.5\textwidth]{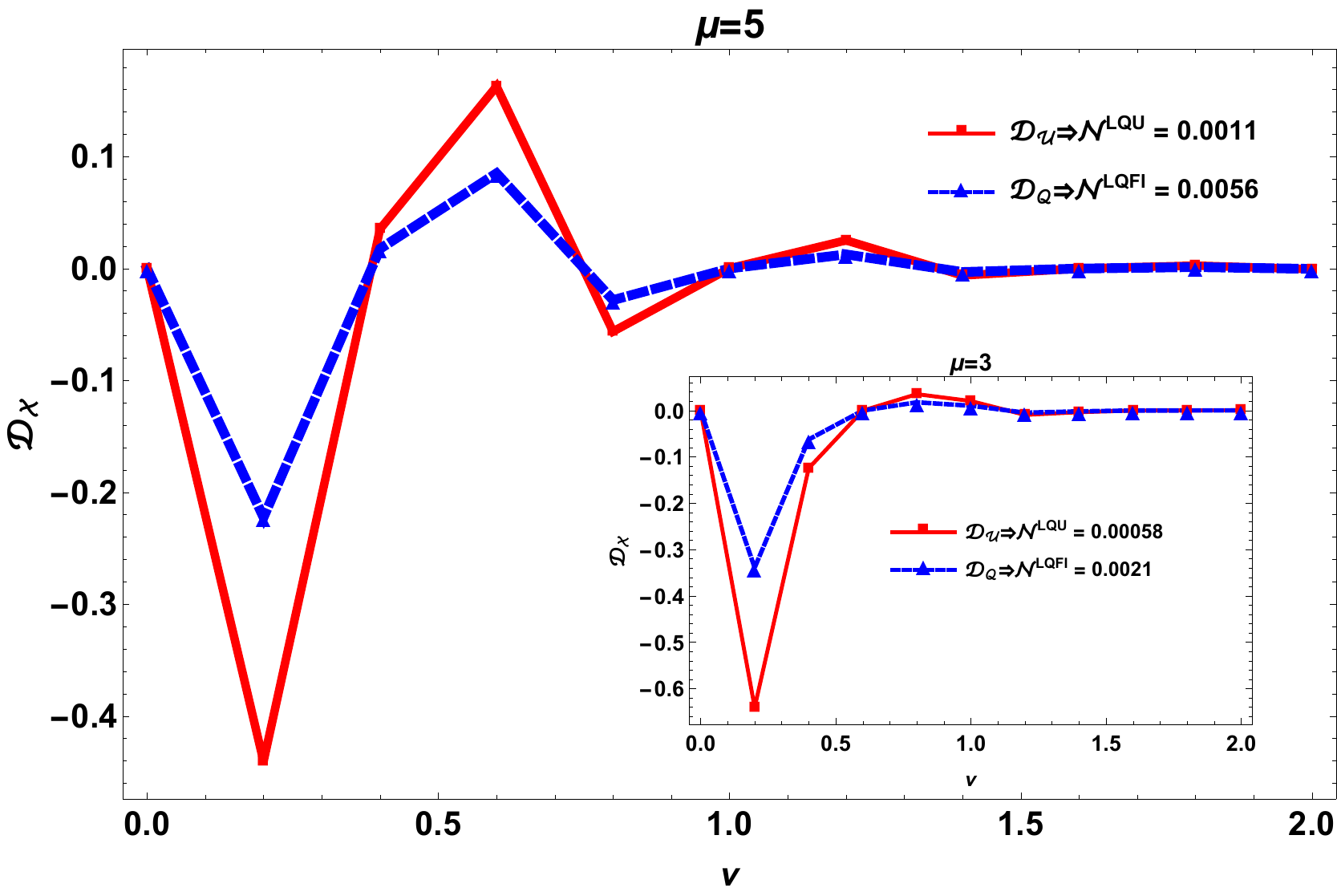}
		\caption{The time derivation $\mathcal{D}_{\mathcal{X}}\left(\mathcal{X}\equiv\mathcal{U,Q}\right)$ versus the dimensionless time $\nu$ for the parameter $\mu=3$ and $\mu=5$. The non-Markovian time periods are those where $D_{\mathcal{X}}>0$.}
		\label{fig:dx35}
	\end{figure}
Here, we have done the same for a non-Markovian depolarization channel based on a different initial state (\ref{rho0}). We obtained the results represented in Fig. (\ref{fig:dx35}) which represents $\mathcal{D}_{\mathcal{X}}\left( \nu\right)>0 \left( \mathcal{X}\equiv \mathcal{U,Q}\right)$ as function $\nu$ for different values of $\mu$. It is worth noting that an increase in the parameter $\mu$ corresponds to an increase in the degree of non-Markovianity, quantified by the measure $\mathcal{N}^{LQFI}$. Conversely, an increase in $\mu$ is associated with a decrease in the degree of non-Markovianity, measured by $\mathcal{N}^{LQU}$. Indeed, as previously the same remarks when $\mathcal{D}_{\mathcal{X}}\left( \nu\right)>0$ mean that the dynamics of the system are non-Markovian, and $\mathcal{D}_{\mathcal{X}}\left( \nu\right)<0$ the system is Markovian. Considering the results obtained here i.e. for a non-Markovian depolarization channel, they agree with the result obtained in the first and second sections for the phase damping channel and the amplitude damping channel means that always $\mathcal{N}^{LQFI}\geq\mathcal{N}^{LQU}$.\par
	
Hence, it is evident from the above examples that the structure of the reservoir influences the non-Markovian nature of the dynamics. In the phase damping dynamical model, non-Markovianity is observed in the super ohmic regime. For the amplitude damping channel, non-Markovianity is present within the small range $0.18 \leq \lambda/\gamma_{0} \leq 0.43$. The dephasing model $\mathcal{N}^{LQFI}$ demonstrates an increasing trend with higher values of $\mu$, suggesting a growing degree of non-Markovianity. Conversely, $\mathcal{N}^{LQU}$ exhibits a decreasing trend with higher values of $\mu$, indicating a reduction in non-Markovian behavior. Indeed, all three examples support the inequality \ref{egalite}.
\section{Concluding Remarks and Outlook}\label{sec4}
Finally, we have developed a new metric for non-Markovianity from the perspective of non-classical correlations using LQFI. This measure has been applied to three common noisy channels, "phase damping channel", "amplitude damping channel" and "depolarising channel". We have carried out a comparative analysis of the relationship between our proposed measure and another measure based on LQU. It has been shown that for these three channels, the proposed measure is consistent with the LQU-based measure.  However, there is a difference in the degree of non-Markovianity, with the LQFI-based measure indicating a greater degree of non-Markovianity compared to the LQU-based measure. Most importantly, the current research suggests that the LQFI-based measure and the LQU-based measure for quantifying non-Markovianity have similar patterns of variation. This work highlights the importance of these specific types of non-Markovianity measures.\par
	
In addition, it is important to highlight the relationships between our measure and other established measures of non-Markovianity, such as information backflow and divisibility. Despite the consistency in detecting non-Markovianity across different measures, our approach is conceptually distinct. For example, the information backflow measure is based on trace distance, while the divisibility measure is based on violations of the divisibility law and the Jamiolkowski-Choi isomorphism. Compared to the LQU-based non-Markovianity measure, our LQFI-based measure offers a unique perspective for capturing the information dynamics of open quantum systems. Our results show that a positive time derivative of LQFI signals the flow of information from the environment to the system, indicating non-Markovian dynamics. This highlights the influence of the reservoir structure on the Markovian/non-Markovian character of an open quantum system. We believe that our measure provides a valuable tool for characterising non-Markovianity, facilitating both theoretical descriptions and experimental investigations. As a truly quantitative measure, it assesses the degree of non-Markovianity in different physical systems and is computationally easier to evaluate compared to LQU-based measures. We expect that non-Markovianity based on LQFI will contribute significantly to further research in open quantum systems. The relationship between different non-Markovian quantifications remains a pertinent topic for future research, despite numerous discoveries in recent years.\par

Building on these results, it is interesting to investigate and explain the experimental feasibility of detecting the CPTP non-Markovianity using LQFI. Its link with measurable quantities in quantum metrology and its precise relationship with quantum correlations are advantageous. This will enable us to recognize the limitations of optimization-based measurements and explore the avenues described above, including the potential role of LQFI and the need for alternative, more experimentally suitable approaches. Addressing this open question is crucial to bridge the gap between theoretical understanding and practical applications of non-Markovian quantum information processing.
	
\end{document}